\documentclass[aps,prd,preprint]{revtex4-1}
\usepackage{graphicx}
\usepackage{subfigure}
\usepackage{caption}
\usepackage{bm}

\newcommand{\Y}{\Y_{lm}}

\newcommand{\be}{\begin{equation}}
\newcommand{\ee}{\end{equation}}
\newcommand{\Be}{\begin{eqnarray}}
\newcommand{\Ee}{\end{eqnarray}}

\newcommand{\f}{\frac}
\begin{document}
\pagestyle{plain}

\title{Evolution of Cosmological Horizons of Wormhole Cosmology}

\author{Sung-Won Kim}
\email[email:]{sungwon@ewha.ac.kr}
\affiliation{Department of Science Education, Ewha Womans University, Seoul 03760, Korea}

\date{today}

\begin{abstract}
Recently we solved the Einstein's field equations to obtain the exact solution of the cosmological model with the Morris-Thorne type wormhole. We found the apparent horizons and analyzed their geometric natures, including the causal structures. We also derived the Hawking temperature near the apparent cosmological horizon.
In this paper, we investigate the dynamic properties of the apparent horizons under the matter-dominated universe and lambda-dominated universe. As a more realistic universe, we also adopt the $\Lambda$CDM universe which contains both the matter and lambda.
The past light cone and the particle horizon is examined for what happens in the case of the model with wormhole. Since the spatial coordinates of the spacetime with the wormhole are limited outside the throat, the past light cone can be operated by removing the smaller-than-wormhole region. The past light cones without wormhole begin start earlier than the past light cones with wormhole in conformal time-proper distance coordinates. The light cone consists of two parts: the information from our universe and the information from other universe or far distant region through the wormhole. Therefore, the particle horizon distance determined from the observer's past light cone can not be defined in a unique way.

\end{abstract}
\maketitle

\section{introduction}
Recently \cite{kim2018} we derived the exact solution of the cosmological model with a wormhole
by using the method
similar to that of McVittie \cite{MV} who had obtained the black hole solution
in the flat Friedmann-Lemaitre-Robertson-Walker (FLRW) universe.
Even though there are several example models for wormhole cosmology \cite{3}-\cite{BF}, they did not
guarantee the Einstein equation.

It is well known that there are two prototypes of wormhole: the Morris-Thorne wormhole \cite{MT} and
the Visser wormhole \cite{visser}.
Thus, the wormhole cosmological model is based on these two models.
Kim \cite{3}, Hochberg \cite{hochberg}, Li \cite{Li}, Bochiccio and Faraoni \cite{BF} dealt with the Visser type thin-shell wormhole to construct a cosmological model with a wormhole.
Usually, we prepare two copies of the appropriate model, cut off the central parts of the copies, and paste the remaining two pieces to the cut edge. This junction plays the role of wormhole throat connecting two universe models. As an example of minimizing the use of exotic matter at the junction,
we can consider the delta function distribution of the matter.
It is an easy way to do it by hands, but it is an unnatural way to create a wormhole.

Another way to construct a cosmological model with a wormhole is to start with a
Morris-Thorne type wormhole. We combine the wormhole and the cosmological background by hand
with plausibility.
Later we can find the satisfactory matter conditions from the Einstein's equation,
but this is not a natural way either.
Roman \cite{roman}, Kim \cite{kim}, Gonz\'{a}lez-D\'{\i}az \cite{Gon}, and Mirza et al. \cite{MED}, used MT wormhole to construct a cosmological models with a wormhole.

The solution  we got \cite{kim2018} was unique in the sense of satisfying Einstein's equation with
Morris-Thorne type wormhole starting from the generic isotropic spacetime.
We also got two apparent horizons by transforming to the Schwarzschild-like
coordinates: one is the apparent cosmological horizon and the other one is the wormhole throat.
It is important to know how the cosmic matter affects the spacetime of the wormhole cosmology and its
horizons,
because we can get the role of wormhole in the universe from the information.
The dynamic nature of the spacetime and cosmological horizons is as important as the causal nature
in analyzing the wormhole cosmological model and understanding the physics of the universe.

In this paper, we considered their dynamic nature of the apparent horizons under the background cosmological models. The background models considered here are the matter- and/or lambda-dominated universe.
Recent observations \cite{planck} reveal that the spacetime is nearly flat ($\Omega_k \simeq 0$) and
the most probable model is $\Lambda$CDM universe. In this case radiation is usually neglected, so the model
with the matter and lambda is simpler.

We need to know how the wormhole affects on the geometry of the universe.
Therefore, in each universe, the creation and dynamics of the apparent horizons, past light cone, and
particle horizon distance are investigated and the characteristics of the
model with the wormhole are found. The apparent cosmological horizon has
the meaning of the modified Hubble horizon, because it approaches the
Hubble horizon in the absence of a wormhole.
The role of the Hubble horizon in the universe is the boundary between particle at a speed
smaller than the speed of light $c$ and the particle at a speed higher than $c$ relative to the observer
of a given time.
Thus we can say that the apparent cosmological horizon constitute a sphere containing
all the astronomical systems, in this spacetime, at the time of observation receding
from the observer slower than the light.
The speeds of the horizons are also discussed because they are useful for viewing their dynamics.
The apparent horizons' speeds are compared with the Hubble surface speed and the wormhole speed.

Since the particle horizon is defined as the maximum distance that particles could have
traveled to the current observer,
this consideration gives us the information of the spacetime's causal structure
as well as the light cone.
The particle horizon distance is decided by the observer's past light cone \cite{Harrison}.
An observer's particle horizon divides all world lines into two classes at the moment of observation:
those that intersect the observer's past light cones, and those that lie outside the reach of the
observer's light cone.
We should know how past light cone and particle horizon distance are changed for spacetime of the
universe with a wormhole.
We can also discussed on the issue of horizon problem
and introduction of the inflation in the early universe in big bang scenario.

In Sec 2, we briefly summary the solution of the wormhole cosmology and the structure of the apparent horizons.
In Sec 3, we investigate the time evolution of the Hubble horizon, particle horizon, light cone for
the model with a wormhole depending on the matters in the universe.
Summary and discussion are in final section.

\section{Exact solution and apparent horizons}

The Morris-Thorne type wormhole (MT-wormhole) is given by \cite{MT}
\be
ds^2 = - e^{2\Phi} dt^2 +   \f{1}{1-b(r)/r} dr^2 + r^2 d\Omega^2,   \label{mt_worm}
\ee
where $\Phi(r)$ is red-shift function and $b(r)$  is the shape function. The geometric unit, that is,
$G=c=\hbar=1$ is used here. The radius $r$ is in the range of $b<r<\infty$. Two functions $\Phi(r)$
and $b(r)$ are restricted by `flare-out condition' to maintain the shape of the wormhole.

The isotropic form of the static wormhole, to see it under FLRW universe, is
\be
ds^2 = - d\tilde{t}^2 + \left( 1 + \f{b_0^2}{4\tilde{r}^2} \right)^2 ( d\tilde{r}^2 + \tilde{r}^2 d\Omega^2 )
\label{isotropic}
\ee
for the ultra-static case ($\Phi'=0,~~ b(r)=b_0^2/r$).
Here coordinates with tilde mean the transformed coordinate to the isotropic wormhole. The radial
coordinate is $\tilde{r} > b_0/2$ in isotropic form (\ref{isotropic}), while $r>b_0$ in spherically
symmetric form
(\ref{mt_worm}).
However, we will omit the tilde over coordinates by convenience from now on, if only there are no confusions.

The FLRW spacetime in isotropic form is given by \cite{MV}
\be
ds^2 = - dt^2 + \f{a^2(t)}{(1+kr^2)^2} ( dr^2 + r^2 d\Omega^2 ).
\ee
Here $a(t)$ is the scale factor and $k=1/4{\cal R}^2$, where ${\cal R}$ is the curvature, and $k$ goes zero in case of flat FLRW spacetime.  We start from the general isotropic metric element to see the unified wormhole in FLRW cosmological model as
\be
ds^2 = - e^{\zeta(r,t)} dt^2 + e^{\nu(r,t)} ( dr^2 + r^2 d\Omega^2 ).
\ee

Einstein's equation is given by
\[
G_{\alpha\beta} = \kappa T_{\alpha\beta},
\]
where $\kappa = 8\pi$.
The non-zero components of Einstein tensor ${G^\mu}_\nu$ are
\Be
{G^0}_0 &=&  \f{1}{4r} \{ [ (8\nu' + 4\nu'' r + \nu'^2 r ) e^{-\nu+\zeta} - 3 \dot{\nu}^2 r ] e^{-\zeta} \}, \\
{G^0}_1 &=& \f{1}{2}(2\dot{\nu}' - \dot{\nu}\zeta') e^{-\zeta},\\
{G^1}_1 &=&  \f{1}{2r}\{ [r(-2\ddot{\nu} + ( -\f{3}{2}\dot{\nu} + \dot{\zeta}) \dot{\nu} ) e^{-\zeta+\nu} + 2\nu' + 2\zeta' + \zeta'\nu'r + \f{1}{2}r \nu'^2 ] e^{-\nu} \},\\
{G^1}_0 &=& \f{1}{2}(-2\dot{\nu}' + \dot{\nu}\zeta') e^{-\nu},\\
{G^2}_2 &=& {G^3}_3 = \f{1}{4r} \{ [  2\zeta' +2\nu' + 2\nu''r + 2\zeta''r + \zeta'^2 r] e^{-\nu}
+2r( -2\ddot{\nu} -\f{3}{2}\dot{\nu}^2 + \dot{\zeta}\dot{\nu} ) e^{-\zeta+\nu} \}.
\Ee
Here dot denotes the derivation with respect to $t$ and prime denotes the derivative with respect to $r$.

By solving the Einstein's equation, the exact cosmological wormhole solution is given by \cite{kim2018}
\be
ds^2 = - dt^2 + \f{a^2(t)}{( kr^2 + 1)^2} \left( 1 + \f{b_0^2}{4r^2} \right)^2 (dr^2 + r^2 d\Omega^2).
\label{worm-cos}
\ee
It is assumed that the radial accretion is not allowed in this model like the Mc Vittie solution \cite{MV}.
Here $a(t)$ is the scale factor and $k$ is the curvature parameter of FLRW cosmology.
The model shows that the cosmic part and the wormhole part are multiplied, so that
the spatial and the temporal parts are separated. The resultant spacetime is natural, since the ansatz
for the matter distribution (cosmic matter and wormhole matter) is separable during finding the solution,
such as
\Be
a^2(t)\rho(r,t) &=& a^2(t)\rho_c(t) + \rho_w(r), \\
a^2(t)p_1(r,t) &=& a^2(t)p_{1c}(t) + p_{1w}(r), \\
a^2(t)p_2(r,t) &=& a^2(t)p_{2c}(t) + p_{2w}(r), \\
a^2(t)p_3(r,t) &=& a^2(t)p_{3c}(t) + p_{3w}(r),
\label{matters}
\Ee
which is the method like the separation of variables in partial differential equation of many physical problems. This ansatz were already used in the previous wormhole cosmological model \cite{kim}.

When we change (\ref{worm-cos}) into the Schwarzschild-like form in new coordinate $R$, the spacetime will be
\be
ds^2 = -\left( 1- \f{R^2H^2}{r^2J^2} \right) dt^2 + \f{1}{r^2J^2} dR^2
- \f{2HR}{r^2J^2} dtdR + R^2 d\Omega^2,
\ee
where
\be
R \equiv a(t) \left(\f{(1+b_0^2/4r^2)}{(1+kr^2)} \right)r = a(t)A(r) \quad \mbox{and} \quad J \equiv \f{A'}{A}.
\ee
To remove $dtdR$-term, by defining
\be
dT = \f{1}{F}(dt+\beta dR),
\ee
the spacetime can be
\be
ds^2 = - \f{H^2({R_+}^2-R^2)(R^2-{R_-}^2)}{R^2-a^2b_0^2}F^2 dT^2
 + \f{1}{r^2J^2}\f{R^2-a^2b_0^2}{H^2({R_+}^2-R^2)(R^2-{R_-}^2)}dR^2 + R^2 d\Omega^2.
\ee
Here $R_\pm$ are the location of the apparent horizons
\be
R_\pm = \f{1}{\sqrt{2H^2}} [  1 \pm \sqrt{1-(2b(t)H)^2}]^{1/2} < R_{\rm H} \equiv \f{1}{H}, \label{horizons}
\ee
where $b(t)\equiv a(t)b_0$ is the extended wormhole with a scale factor, as seen in the isotropic metric of the wormhole.
This is the previous time dependencies of the cosmological wormholes \cite{roman,kim}
that expand along the scale factor. The modified wormhole in their works also satisfied the
flare-out condition and their roles in the spacetime.
Here, $R_+$ means the apparent cosmological horizon, and $R_-$ is the minimum size of dynamic wormhole throat
in this spacetime. The condition of $(2b(t)H)^2<1$ is required for these two horizons $R_\pm$
to be real values.
This condition decide the existence of the two horizons and the coincidence time of them.

For the FLRW universe without wormholes, the Hubble surface defined by $R_{\rm H}=1/H$ that is the boundary between the subluminal inner sphere and the superluminal outer sphere.
The Hubble horizon $R_{\rm H}$
%is the apparent cosmological horizon of the FLRW model without wormhole and
is always outside the apparent horizons. If the wormhole vanishes in the universe, $b_0=0$, $R_+$ approaches the Hubble horizon and $R_-$ disappears. Thus, the apparent cosmological horizon acts as the Hubble horizon in the universe with wormhole. In the region smaller than the sphere of radius $R_+$ of this spacetime, any particle travels slower than the light.
At the limit of $a(t)$=const ($H=0$), $R_-$ approaches to $b_0$.
That is, when the cosmological background is removed from this model, the spacetime becomes the MT wormhole.

\section{Time evolution of the horizon}

\subsection{Power law distribution}

Now we apply the issues to the various distributions for cosmic part of the universe.
Above all, we have to check the Friedmann equation to see if there are changes
in the equation by a wormhole in this spacetime.
The Friedmann equation for the wormhole cosmological model is
\be
H^2 \equiv  \left( \f{\dot{a}}{a} \right)^2 = \f{8\pi \rho_c}{3},
\ee
which is the same as for a model without a wormhole because time-dependence
of $\rho_c$ and time-independence of $\rho_w$ are completely separated as wee see in (\ref{matters}).
With energy conservation law
$\dot{\rho}_c + 3H(\rho_c + P_c ) = 0$
and the equation of state $P_c = \omega \rho_c$, the scale factor $a$ and Hubble parameter will be
\be
a = \left( \f{t}{t_0} \right)^{2/(3+3\omega)} \equiv \left( \f{t}{t_0} \right)^q \quad
\mbox{and}  \quad
H \equiv \f{\dot{a}}{a} = \f{q}{t}, \label{time-dep}
\ee
when $a(t)$ follows the power law and $\omega$ is the parameter of the equation of state.
The present time, the age of our universe $t_0$ is
\be
t_0 = \f{1}{1+\omega}\left( \f{1}{6\pi\rho_c} \right)^{1/2} \quad (\omega \neq -1)
\ee
and $a_0\equiv a(t_0)=1$, is the present scale factor.
The Hubble surface recede at the radial velocity
\be
\f{dR_{\rm H}}{dt} = \f{1}{q} = 1 + \xi,
\ee
where $\xi = - \ddot{a}a/\dot{a}^2$ the deceleration parameter and the dot means the differentiation with
respect to time $t$.

Now we discuss the time evolutions of apparent horizons according to the cosmic matter distributions
of the background universe. Since the background universe model is FLRW, we will consider the matter-universe models with and/or without lambda. Due to the time-dependence
$a(t)$ and $H(t)$, the apparent horizons do evolve and expand or collapse.
The apparent cosmological horizons $R_\pm$ with time-dependencies (\ref{time-dep}) for the power law, are represented in terms of time $t$ as
\be
R_\pm(t) = \f{t}{\sqrt{2}|q|}[1 \pm \sqrt{1-(2qb_r)^2(t/t_0)^{2(q-1)}}]^{1/2}, \label{horizon_time}
\ee
where $b_r\equiv  b_0/t_0$ is the relative wormhole size. The time when two horizons coincide is
\be
t_{\rm c} = ( 2qb_r)^{1/(1-q)}t_0
\ee
and the size of the resultant horizons is
\be
R_{\rm c} = \f{(2qb_r)^{1/(1-q)}}{\sqrt{2q^2}} t_0  = \f{t_{\rm c}}{\sqrt{2q^2}}
\ee
at that time. %The coincidence time is earlier as the value of $q$ is smaller.
The wormhole has the size restricted as
\be
b_r < \f{1}{2q}\left(\f{t}{t_0}\right)^{(1-q)}
\ee
for the existence of $R_\pm$ in (\ref{horizon_time}).

To see the time evolutions of the horizon, the time derivatives of the horizons are
\Be
 \f{d}{dt}R_\pm &=& \f{1}{\sqrt{2}|q|}
\left\{ [1\pm \sqrt{1-Bt^{2(q-1)}}]^{1/2} \right. \nonumber \\
&& \left. \mp \f{1}{2}
\f{B t^{2(q-1)}(q-1)}{[1 \pm \sqrt{1-Bt^{2(q-1)}}]^{1/2}\sqrt{1-Bt^{2(q-1)}}}
\right\}, \label{time_dependence}
\Ee
where $B = (2qb_0)^2(t_0)^{-2(q-1)}$.
The $R_-$ has the minimum value  at
\be
t_{\rm min} = [\sqrt{q}(1+q) b_r ]^{1/(1-q)}t_0
\ee
just after the coincidence time $t_{\rm c}$ and increases after $t_{\rm min}$ with (\ref{time_dependence}).
The ratio of $t_{\rm min}$ to $t_{\rm c}$ is
\be
\f{t_{\rm min}}{t_{\rm c}} = \left( \f{(1+q)^2}{4q} \right)^{1/2(1-q)} > 1, \quad \mbox{for} \quad 0<q<1
\ee
and is depends only on $q$ and does not depend on $b_r$ or $t_0$.
This ratio is larger as the value of $q$ is smaller.
The minimum value of $R_-$ at $t=t_{\rm min}$ is
\be
R_{-}^{(\rm min)} = \f{1}{\sqrt{q(1+q)}} \f{t_{\rm min}}{t_0}.
\ee
The reason of the existence of minimum value of $R_-$ is
that $R_-$ is given as the product of the linearly increasing part
and the decreasing part (for $q<1$) in time as shown in (\ref{horizon_time}). Therefore,
$\dot{R}_-<0$ before $t_{\rm min}$ and $\dot{R}_->0$ after $t_{\rm min}$.
It can be said that the wormhole is initially contracted and
expands with the very high expansion of the universe.
The expansion of the universe close to the coincidence time prevails
the expansion of the wormhole, so the wormhole shrinks before $t_{\rm min}$.

At very early times when the horizons was created, the speeds of the apparent
horizons diverge near $t_{\rm c}$, such as
\be
\lim_{t \rightarrow t_{\rm c}} \f{dR_{\pm}}{dt} = \pm \infty.
\ee
As time goes on, $dR_+/dt$ decreases and $dR_-/dt$ increases from negative to positive,
so $R_-$ has the minimum value.
In the distant future, the speeds of the horizons are given as
\Be
\lim_{t\rightarrow\infty} \f{dR_+}{dt} &=& \f{1}{|q|}, \\
\lim_{t\rightarrow\infty} \f{dR_-}{dt} &=& b_0q\left(\f{t}{t_0}\right)^{(q-1)}=b_0\dot{a},
\Ee
for positive $q$. The speed of ${R}_+$ approaches the speed of Hubble surface and $R_-$ will
eventually stop when $q<1$ in flat universe.

\subsection{Matter-dominated universe}

%First, we see a single component universe as a simple example of our cosmological model.
%Now we consider the three cases of the single component universe, the matter-dominated universe %$(\omega=0)$,
%the radiation-dominated universe $(\omega=1/3)$, and the lambda-dominated universe $(\omega=-1)$.

For matter-dominated universe $\omega=0$, $q=2/3$, $a(t)=(t/t_0)^{2/3},$ and $H(t)=\f{2}{3t}$. The time-dependencies of
apparent horizons and their speeds are depicted in Fig. 1. The equation (8) shows that two horizons
do not appear before $t_{\rm c}^{\rm (m)}$, where $t_{\rm c}^{\rm (m)}=(4b_r/3)^3t_0$ is the time when the
two horizons are generated and separated simultaneously in the matter-dominated universe.
Prior to $t_{\rm c}^{\rm (m)}$, the wormhole matters are highly localized at very tiny region and
cannot generate $R_\pm$, because $H$ is very large as can be seen in the square root of the equation
(\ref{horizon_time}). Then, as the speed of the scale factor decreases, the horizons $R_\pm$ appear and
expand with the scale factor.
The size of the horizons at the coincidence time is just
\be
R_{\rm c}^{\rm (m)}= \f{16}{9}\sqrt{2}b_r^3 t_0.
\ee
For larger the wormhole, the later the coincidence time. When the more matter is concentrated to
construct the wormhole, the generation time of wormhole and apparent horizon will be delayed.
When universe expands, the two horizons also expand, but the expansion rate of $R_+$ is higher than that
of $R_-$, so they can never meet again. At a late time, $R_+$ approaches $R_{\rm H}$ and $R_-$ approaches
$b(t)=b_0(t/t_0)^{2/3}$. Just after the coincidence time,
the size of the wormhole, $R_-$, decreases before the time 
$t_{\rm min}^{(\rm m)}=(\sqrt{50/27}b_r)^3t_0$ and increases after that time.
The ratio of $t_{\rm min}^{\rm (m)}$ to $t_{\rm c}^{\rm (m)}$ is
\be
\f{t_{\rm min}^{\rm (m)}}{t_{\rm c}^{\rm (m)}} = \left( 1 + \f{1}{24} \right)^{3/2}.
\ee
The minimum size of the wormhole at that time is
\be
R_{-}^{(\rm min)} = \f{50}{81}\sqrt{15}b_r^3.%, \quad \mbox{for} \quad \omega=0.
\ee
If the wormhole size $b_0$ is smaller,  the time of wormhole
minimum size is closer to $t=0$, since $t_{\rm c}$ is proportional to $b_r^3$.

For the speed of horizon as we see Fig. 1(b), the speed of $R_+$ is very large near the coincidence time and
rapidly approach the speed of $R_{\rm H}$. At the very early time, $dR_+/dt > dR_{\rm H}/dt$,
and later the inequality sign reverses. Finally $dR_+/dt$ approaches $dR_{\rm H}/dt$. The speed of $R_-$ approaches
  to $-\infty$ near coincidence time and will change from negative to positive at $t=t_{\rm min}^{(\rm m)}$.
  Finally $dR_-/dt$ approaches $b_0(t_0/t)^{(1/3)}$ and in the far future $R_-$ will stop expanding
  like the scale factor of zero curvature cosmological model.

%\begin{figure}[h]
%\includegraphics[height=6.5cm]{matter-dominated}
%\caption{Time dependence of apparent horizons for matter-dominated universe,
% Here we set $b_0=1$ (red) and %$b_0=0.75$ (blue).
%Solid lines are $R_+$, dashed lines are $R_-$, dotted line (black) is $R_H$ which is asymptote of $R_+$,
%and the dash-dot lines (green) are $b(t)$ which are asymptotes of $R_-$.
%Two horizons appear at $t_{\rm %c}^{\rm (m)}=t_0$ (for $b_0=0.75t_0$) and $(4/3)^3t_0$ (for $b_0=t_0$),
%before which the singularity was %naked. }
%\end{figure}

%\begin{figure}[h]
%\includegraphics[height=6.5cm]{vel_matter}
%\caption{Time dependence of apparent horizons' velocity for matter-dominated universe, Here we set $b_0=1$
%(red) and $b_0=0.75$ (blue).
%Solid lines are $R_+$, dashed lines are $R_-$, dotted line (black) is $R_H$ which is asymptote of $R_+$,
%and the dash-dot lines (green) are $b(t)$ which are asymptotes of $R_-$.
%Two horizons appear at $t_{\rm %c}^{\rm (m)}=t_0$ (for $b_0=0.75t_0$) and $(4/3)^3t_0$ (for $b_0=t_0$),
%before which the singularity was %naked. }
%\end{figure}

\begin{figure*}
  \centering
  \subfigure[Horizons]{%
    \includegraphics[width=0.5\textwidth]{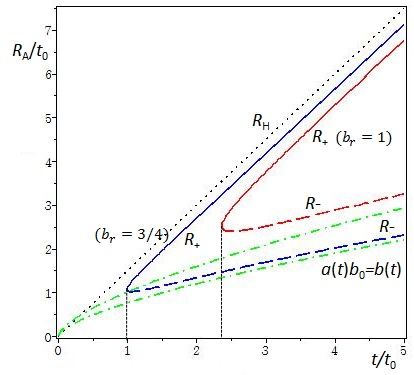}%
    \label{fig:a}%
  }%
  \hfill
  \subfigure[Speed of horizons]{%
    \includegraphics[width=0.45\textwidth]{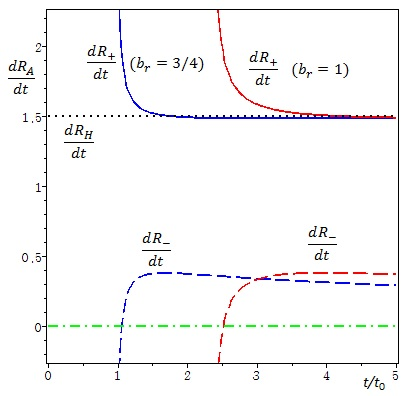}%
    \label{fig:b}%
  }%
  \caption{(a) Time dependencies of apparent horizons in matter-dominated universe. Here we set $b_r=1$ (red)
  and $b_r=0.75$ (blue) close to zero. The solid lines are $R_+$, dashed lines are $R_-$, dotted line
  (black) is $R_{\rm H}$
  which is asymptote of $R_+$, and the dash-dot lines (green) are $b(t)$ which are asymptotes of $R_-$.
  Two horizons appear at $t_{\rm c}^{\rm (m)}=t_0$ (for $b_r=0.75t_0$) and $(4/3)^3t_0$ (for $b_r=1$).
  (b) Time dependencies of apparent horizons' speeds for matter-dominated universe. Here we set
  $b_r=0.75$ (blue)
  and $b_r=1$ (red). Solid lines are speeds of $R_+$, dashed lines are speeds of $R_-$, and
  the dot line is the speed of $R_H$.}
  %\label{fig:ab}
\end{figure*}

\subsection{Lambda-dominated universe}

In the lambda-dominated universe, the attributes of $a, \dot{a}, H$, and $\dot{H}$ are quite different from
those of power-law expansion of the scale factor. In this universe, $a=e^{H_0(t-t_0)}$ instead of the power
law and $H=H_0 = {\rm const}$.  At very early times, they are completely separate 
as shown in Fig 2. Two horizons exist for $t<t^{\rm (\Lambda)}_{\rm c}=t_0-\ln(2H_0t_0b_r)/H_0$ and
two horizons meet at $t^{\rm (\Lambda)}_{\rm c}$. After $t^{\rm (\Lambda)}_{\rm c}$, two horizons disappear
and the wormhole also disappears. Thereafter, the huge exponential growth of the scale factor prevents
the existence of the wormhole and the cosmological horizon.
The size of two horizons at $t^{\rm (\Lambda)}_{\rm c}$ are
\be
R_{\rm c}^{\rm (\Lambda)}=\f{1}{\sqrt{2}}\f{1}{H_0}.
\ee
We can compare this result with the model by Roman \cite{roman}, where the wormhole expands along the scale
factor and exponentially increases. As a result, the wormhole does not disappear in his paper.
 At very early times, $R_+$ approaches $R_H$ and $R_-$ approaches $b_0e^{H_0(t-t_0)}=b(t)$.
This looks like a wormhole model in de Sitter universe. Also, when if you look at their
behaviors of $\dot{R_\pm}$ near the coincidence time, they diverge to $\pm\infty$ and the meeting point is
smooth. The two apparent horizons $R_\pm$ do not appear again after the coincidence time.
The detailed forms of the horizons' speeds are
\be
\f{dR_\pm}{dt} = \mp \sqrt{2} \f{b_0^2 H_0^2 e^{2H_0(t-t_0)}}{[1\pm\sqrt{1-4b_0^2 H_0^2
e^{2H_0(t-t_0)}}]^{1/2}\sqrt{1-4b_0^2 H_0^2 e^{2H_0(t-t_0)}}}. \label{speed}
\ee
As we see in (\ref{speed}), $\dot{R}_+$ is always negative and $\dot{R}_-$ is positive.
It means that $R_+$ always decreases and $R_-$ increases by the coincidence time.

%\begin{figure}[h]
%\includegraphics[height=6cm]{lambda-dominated}
%\caption{Time dependence of apparent horizons for $\Lambda$-dominated universe, Here we set $a_0=b_0=1$ and
%$H_0=0.05$. The time $t_{\rm \Lambda 0}=16$.
%Solid line is $R_+$, dashed line is $R_-$, dotted line is $R_H$ which is asymptote of $R_+$, and
%dash-dotted line is $b(t)$ which is the asymptote of $R_-$.}
%\end{figure}

\begin{figure*}
  \centering
  \subfigure[Horizons]{%
    \includegraphics[width=0.5\textwidth]{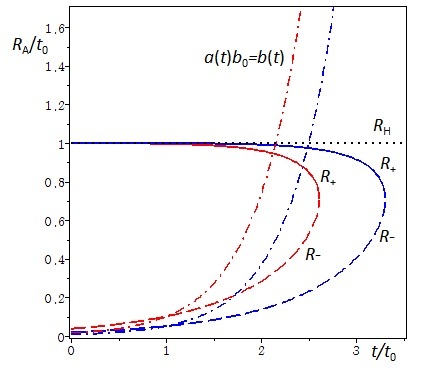}%
    \label{fig:a}%
  }%
  \hfill
  \subfigure[Speed of horizons]{%
    \includegraphics[width=0.45\textwidth]{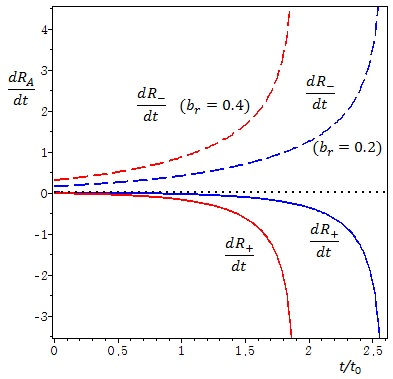}%
    \label{fig:b}%
  }%
  \caption{(a) Time dependencies of apparent horizons in $\Lambda$-dominated universe. Here we set
  $a_0=1$, $b_0=0.2$ (blue), $b_0=0.4$ (red), and $H_0=0.05$. The time $t_{\rm \Lambda 0}=16$.
Solid lines are $R_+$, dashed lines are $R_-$, dotted line is $R_H$ which is asymptote of $R_+$, and
dash-dotted line is $b(t)=b_0e^{H_0(t-t_0)}$ which is the asymptote of $R_-$. (b) Time dependencies
of apparent horizons' speed for lambda-dominated universe with the same values of parameters as (a). }
  \label{fig:ab}
\end{figure*}

\subsection{$\Lambda$CDM Universe}

Here we consider the $\Lambda$CDM model as a more realistic universe, where the dominant components
of the universe are matter and lambda. There are various properties depending on the ratio of matter to
$\Lambda$. The negative lambda makes the universe a closed universe model. In this paper, however,
we will only treat lambda as positive one. The ratio of lambda density to critical
density is given by $\Omega_{\Lambda0} = 1 - \Omega_{\rm m0}$ and the Friedmann equation is
\be
\f{H^2}{H_0^2} = \f{\Omega_{\rm m0}}{a^3} + \Omega_{\rm \Lambda 0}, \label{friedmann_multi}
\ee
where $\Omega_{\rm m0}$ and $\Omega_{\rm \Lambda 0}$ are matter density ratio to the critical density and
lambda density ratios to the critical density at the present time, respectively.
The first term is from the matter-dominated universe and the second term is from
the lambda-dominated universe. This is very similar to the non-zero curvature universe and the second term
here serves as a curvature term.
In this case, since the time dependence is a very complicated form, we will show the related equations
here in terms of redshift $z$ defined by
\be
(1+z) \equiv \f{1}{a}.
\ee
Hubble parameter $H(z)$ is given in (\ref{friedmann_multi}) as
\be
H(z) = H_0 \sqrt{\Omega_{\rm m0}(1+z)^3 + \Omega_{\rm \Lambda 0}}.
\ee
The time at which two horizons coincide is the solution to the equation from the inner square root of the
definition of $R_\pm$ in (\ref{horizons}) as
\be
\Omega_{\rm m0}(1+z)^3 -\left( \f{1}{2b_0H_0} \right)^2 (1+z)^2 + \Omega_{\rm \Lambda 0} = 0.
\ee

This model is a combination of two dominant universes, so there are two coincidence times: the earlier
coincidence time in matter-dominated case and the later coincidence time in lambda-dominated universe.
Hence the coincidence times can be found in two extreme limits.
As wee see in Fig. 3, the later critical time is shown near $z=-1$, while the earlier time is near
$z \rightarrow \infty$ whose case is not shown here. When we extend $z$ infinitely, we will see that two horizons meet.

In the limit of $z+1 \rightarrow \infty$, the solution is approximately
\be
(1+z) \approx \left( \f{1}{2b_0H_0} \right)^2 \f{1}{\Omega_{\rm m0}},
\ee
which is the earlier coincidence time $t_1$ due to $\Omega_{\rm m0}$.
If we represent this time in terms of $t$, we can see that this time is equal to
$t_{\rm c}^{(m)}=(4b_r/3)^3t_0$ of the matter-dominated single-component universe with $\Omega_{\rm m0}=1$.
In the limit of $z+1 \rightarrow 0$, the solution is
\be
(1+z) \approx (2b_0H_0) \sqrt{\Omega_{\rm \Lambda 0}}
\ee
which is the late coincidence time $t_2$ due to $\Omega_{\rm \Lambda 0}$.
If we set $\Omega_{\rm \Lambda 0}=1$, this time is equal to
$t_{\rm c}^{(\Lambda)}=t_0-\ln(2H_0t_0b_r)/H_0$ of the lambda-dominated single-component universe.
The larger the size of the wormhole throat $b_0$, the closer the later meeting time is to $z=-1$.
If the ratio $\Omega_{\rm m0}/\Omega_{\rm \Lambda 0}$ is larger, the coincidence time $t_1$ is earlier.
The smaller the ratio, the later the coincidence time $t_2$.

\begin{figure*}%[h]
\includegraphics[height=6.5cm]{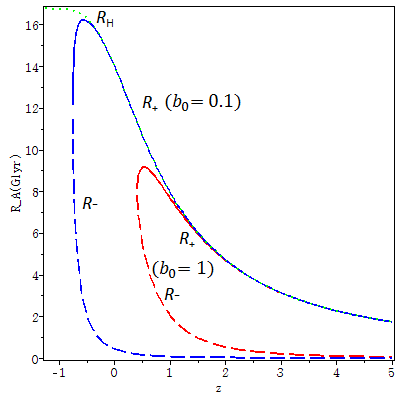}
\caption{Redshift dependence of apparent horizons for matter and $\Lambda$-dominated universe.
Here we set $b_0=0.1$ (blue) and $1$ (red), $H_0=1/14, \Omega_{\rm m0}=0.3, \Omega_{\rm \Lambda 0}=0.7$.
Solid lines are $R_+$, dashed lines are $R_-$, dotted line is $R_H$ which is asymptote of $R_+$.}
\end{figure*}

\subsection{Past light cone and particle horizon distance}

The past light cone is defined as a past null geodesics in the spacetime.
The proper distance in the spacetime of the wormhole cosmological model is
\be
d_p(t) = a(t) \int_0^r \left( 1 + \f{b_0^2}{4r^2} \right) dr = a(t) r \left( 1 - \f{b_0^2}{4r^2} \right) \equiv
a(t)r^*,
\ee
which is shortened to the factor of $( 1 - b_0^2/4r^2 )$ due to wormhole effect
because of the range of the coordinate $r$. The larger wormhole
is subject to a more reduction in space near the origin.
However, the null radial geodesic (constant $\theta, \phi$) is given by
\be
dt^2 = a^2 \left( 1 + \f{b_0^2}{4r^2} \right)^2 dr^2,
\ee
and the conformal time is just the proper distance along the null geodesic as
\be
\int \f{dt}{a} = \int_0^r \left( 1 + \f{b_0^2}{4r^2} \right) dr = r^*.
\ee

As can be seen in Fig.~4, the null dashed line is the past light cone for the universe without wormhole
in $r-\tau$ coordinates spacetime.
However, the past light cone in wormhole cosmology is determined by the different proper time
$r_*=r(1-b_0^2/4r^2)$ instead of $r$ (Fig.4(a)). Thus the line of the null geodesic for wormhole cosmology
is timelike solid line in the figure.
If we eliminate the region $r<b_0/2$ and combine two solid lines, it becomes the past light cone for the
wormhole universe. It is narrower than that in the universe without a wormhole whose null geodesic is in dashed line.

If they are represented together in $r_*-\tau$ coordinates spacetime in Fig.~4(b), the modified past light cone shows the
null line $\tau=r_*$ and past light cone of the case without wormhole shows
the spacelike line $\tau=r=\f{1}{2}[r_*+\sqrt{R_*^2+b_0^2}]$.
The latter (the universe with wormhole) starts earlier by $\tau=b_0/2$ than the former (the universe without wormhole), as shown in Fig.~4(b).
Of course, if we draw the past light cones in $(2+1)$-dimensional spacetime with $r_*-\tau$ coordinates,
they are null cone in the case with wormhole and spacelike cone
in the case without wormhole, as shown in Fig.~4(c).

%\begin{figure}[h]
%\includegraphics[height=5cm]{past_light_cone1}
%\caption{Past null lines in the cases of the universe ``with (solid line)" and ``without (dashed line)"
%wormhole.
%In (a), the spacetime is $r-\tau$ coordinates.
%The modified past null line is outer than $b_0/2$ even at $\tau=0$
%and timelike in this spacetime. The horizontal axis in (b) is $r_*$, the proper distance for universe
%with wormhole. The vertical axis is conformal time coordinate. The past null line is spacelike and starts
%at $\tau=b_0/2$ not origin. (c) A observer's past light cones in the universe with wormhole (solid) and
%without wormhole (dashed) in (2+1) dimensional proper distance-conformal time coordinates.}
%\end{figure}

\begin{figure*}
  \centering
  \subfigure[]{%
    \includegraphics[width=0.30\textwidth]{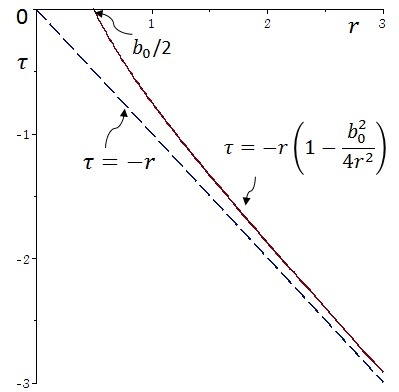}%
    \label{fig:a}%
  }%
  \hfill
  \subfigure[]{%
    \includegraphics[width=0.30\textwidth]{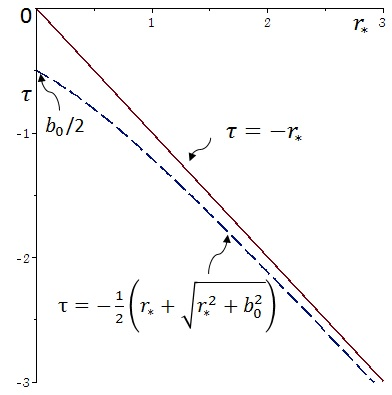}%
    \label{fig:b}%
  }%
  \hfill
  \subfigure[]{%
    \includegraphics[width=0.30\textwidth]{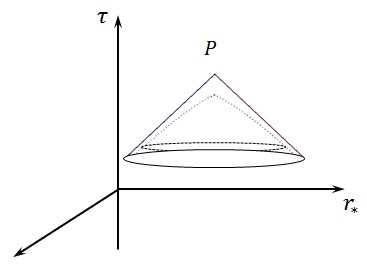}%
    \label{fig:c}%
  }%
  \caption{Past null lines and past light cones in the universe ``with (solid line)" and ``without (dashed line)" wormhole.
(a) The spacetime is in $r-\tau$ coordinates. The past null line for the universe with wormhole at $\tau=0$
is $b_0/2$ more outward than the one for the universe without a wormhole, so it is timelike in this coordinates.
(b) The horizontal axis is $r_*$, the proper distance for universe with a wormhole.
The vertical axis is conformal time coordinate. The past null line for the universe without a
wormhole is spacelike and starts earlier by $\tau=b_0/2$ than the one for the universe with a wormhole.
(c) A observer's past light cones for the universe with a wormhole (solid) and
without a wormhole (dashed) in (2+1) dimensional spacetime in $r_*-\tau$ coordinates.}
 % \label{fig:ab}
\end{figure*}

%\begin{figure}[h]
%\includegraphics[height=5cm]{past_light_cone}
%\caption{past light cone in the cases of the universe ``with (solid line)" and ``without (dotted line)"
%wormhole. The horizontal coordinate is $r_*$, the proper distance for universe with wormhole. The vertical
%axis is conformal time coordinate. When the region $r<b_0/2$ is eliminated, the dotted light cone can be a
%light cone for wormhole universe and it is narrower than the case the universe without wormhole.}
%\end{figure}

 \begin{figure}[h]
\includegraphics[height=6cm]{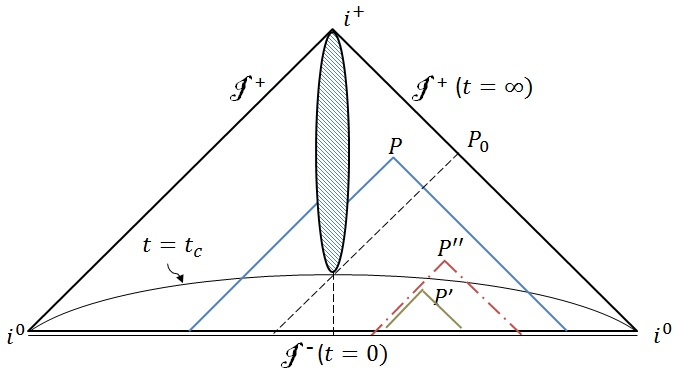}
\caption{Penrose diagram of a wormhole cosmology spacetime with $k=\omega=0$.  The past light cone can be
extended
through the wormhole into the other universe. The particle horizon distance can be also extended.
The observers such as the one earlier than
the generation of the wormhole ($P'$) or the one far from the wormhole ($P''$), its past light cone will not
pass through the wormhole.
However, the observer $P$'s past light cone include the region of the other's universe or the region
far from the observer. The past light cone of the observer $P_0$ is the boundary of the two cases.}
\end{figure}

A Penrose diagram of the wormhole cosmological model is shown in Fig. 5.
It is the just the case of curvature zero and matter-dominated universe.
At the coordinate origin, there is a dynamical wormhole that starts at a time later than
the initial singularity. The spacelike surface of coincidence time $t=t_{\rm c}$ is drawn in figure.
We can think of the past light cones for several observers.
When the observer such as $P'$ is earlier than
the generation of the wormhole, or the observer
such as $P''$ is far from the wormhole, their past light cones will not pass through the wormhole.
However, the past light cone of the observer such as $P$ who exists after the creation
and near the wormhole, include a region of the other's universe or a region
far from the observer, %because $P$'s past information can travel through the throat of the wormhole.
because the past light cone of $P$ consists of the information from the same universe and the information
from the other universe through the throat of the wormhole.
Therefore, the past light cone after the time $t=t_{\rm c}$ is not defined unique way,
in case of the travel-through-wormhole at least. As shown in the figure, the past light cone can be
extended into another universe or far distant causally disconnected regions through the wormhole.
The past light cone of the observer $P_0$ is the boundary between the case where it can pass through
the wormhole and the case where it can not.

The particle horizon distance is determined from the distance from the observer to the end
of the past light cone \cite{Harrison}. In the following three cases there are problems with
the existence of the particle horizon:
(1) When the universe originates in the infinite past in conformal time, particle horizon does not
exist \cite{Harrison}.
(2) In the case of the accelerating (de Sitter steady state, and $q>1$) universe of continual
accelerated expansion,
particle horizon does not exist  \cite{Harrison}.
(3) In the universe with non-trivial topology such as wormhole cosmology, the distance of the past light cone
is not unique. Thus the particle horizon distance is not determined in unique way
for this universe with a wormhole.
The distance of the past light cone from the observer to other universe
by travel through the wormhole is quite different from the one in the same universe.
When we think of a model with wormhole at the very early times, the two causally disconnected regions
can be communicated with each other through the throat of the wormhole.
This means that the horizon problem did not occur.
This result is in accord with the Hochberg's model \cite{hochberg}.

For the universe of constant $\xi$, deceleration parameter, with no wormholes,
the particle horizon distance is $R_{\rm p} = R_{\rm H}/\xi$
and the speed of it is $dR_{\rm p}/dt = 1 + 1/\xi$ in (16).
Thus $R_{\rm p} > R_{\rm H}$ for matter-dominated universe and
$R_{\rm p} = R_{\rm H}$ for radiation-dominated universe.
However, in the case of wormhole cosmological model, the proper distance to the particle horizon is
reduced by wormhole factor $(1-b_0^2/4r^2)$, after the generation of the wormhole,
and the Hubble horizon is replaced by
the apparent cosmological horizon.
The effect of the wormhole factor on the proper distance is dominant near the wormhole,
but there is a travel through the wormhole throat  like the observer $P$ in Fig. 5.
When the observer is far from the wormhole such as $P''$, the observer's
past light cone does not pass the wormhole, and the particle horizon is still larger than the modified Hubble
horizon $R_+$.
Therefore, we can say that the presence of a wormhole in our universe does not change the relationship
between the particle horizon and the Hubble horizon,
if only the observer is far from the wormhole sufficiently enough prevent from
the travel through the wormhole throat.

\section{summary and discussions}
In this paper, we investigated the cosmological horizon of the wormhole cosmological model
for matter-dominant universe and/or lambda.
We checked the properties of the time dependencies of the apparent horizons and the role of wormhole in the universe.
After the wormhole is created at Planck time with Planck size, the wormhole also expands along the
scale factor. The apparent cosmological horizon, the modified Hubble horizon has the meaning of
a boundary between observable and unobservable regions of the universe.

The observer's past light cone depends on proper distance in the spacetime.
The past light cones are examined for what happen in the model with a wormhole.
Because the smaller-than-wormhole region is eliminated from the spacetime for the universe with a wormhole, the past light cone is narrower than the one in the universe without a wormhole.
So the past light cone in the universe without a wormhole starts earlier than
that in the universe with a wormhole in proper distance coordinates.
Moreover, the travel-through-wormhole throat can shows a different way of communication
from normal travel.
Therefore, the particle horizon distance determined from the observer's past light cone can not be
defined unique way.
For the case of the travel through the wormhole, the past light cone can be extended into another universe or far distant causally disconnected region.
In this way we can say that the horizon problem does not occur in this wormhole cosmological model,
because there were causal contact between two
separated regions through wormhole mouth at very early times.
When the observer is far from the wormhole or before generation of the wormhole,
the particle horizon distance is the same as the wormhole-less case
except that the proper distance is shortened by the wormhole factor after the generation of the wormhole.

We can think about the stability issue of the wormhole in the universe.
As shown in the dynamics of the apparent horizons, the size of the wormhole throat
varies with time and expands according to the scale factor. That is, the wormhole is unstable,
even though our model does not allow the inflow.
Such result have already been covered by Shinkai and Hayward \cite{Shinkai}, which showed that
a wormhole will collapse to black hole or expand to a huge size due to a normal or
ghost matter inflows.

Another issue is the creation of a wormhole in the early time of the universe.
Usually we consider the generation of wormholes in the early time, taking into
consideration the quantum era and foams caused by the fluctuation of spacetime.
In our model, instead of this, as we see,
the construction of the wormhole was delayed by a very short time after big bang.
In other words, the expansion rate of the scale factor is very fast, so a highly localized exotic matter
for constructing a wormhole could not form any wormhole at the big bang time.
Later, however, when the expansion rate of the universe falls down in a very short time,
the wormhole and apparent cosmological horizons are created and expand along the scale factor.
Even though there might be a question about the destruction of the wormhole structure due to
the interaction between the background matter and the wormhole matter, the highly localized exotic
matter maintains the wormhole and rapidly expands with
scale factor without mixing with the background matter.

\begin{acknowledgments}
This work was supported by National Research Foundation of Korea (NRF) funded by the Ministry of
Education (2017-R1D1A1B03031081).

\end{acknowledgments}

%\bibliography{apssamp}

\end{document}